\documentstyle[11pt,paspconf,epsf]{article}

\begin{document}

\title{BLR Modeling: A New Approach}
\author{Shai Kaspi and Hagai Netzer}
\affil{School of Physics and Astronomy and the Wise Observatory,
The Raymond and Beverly Sackler Faculty of Exact Sciences,
Tel-Aviv University, Tel-Aviv 69978, Israel}

\begin{abstract}

We present a new scheme for modeling the broad line region in active
galactic nuclei. It involves photoionization calculations applied to a
number of variable emission lines at {\it all times}. We demonstrate how
fitting all lines simultaneously provide strong constraints on several
of the more important parameters, such as the density and column density,
and the radial distribution of the emission line clouds.

When applying the model to the Seyfert 1 galaxy NGC~5548, we are able
to reconstruct the light curves of four emission-lines, in time and in
absolute flux. We argue that the Balmer line light curves, and possibly
also the Mg\,{\sc ii}$\lambda$2798\AA\ light curve, do not fit this
scheme because of the limitations of present-day photoionization
codes. We rule out models where the particle density scales as
$r^{-2}$ and favor models where it scales as $r^{-(1-1.5)}$. We
can place lower limits on the column density at a distance of 10
ld, of $N_{col}(r$=10)${\mathrel{\raise.3ex\hbox{$>$}\mkern-14mu
\lower0.6ex\hbox{$\sim$}}} 10^{22}$ cm$^{-2}$, and limit the particle
density to be in the range of 10$^{11}$$>$$N(r$=10)$>$10$^{9.5}$
cm$^{-3}$.

\end{abstract}


\section{Introduction}

Broad emission line regions (BLRs) in Active Galactic Nuclei (AGNs)
have been the subject of extensive studies for more than two decades.
Such regions are not spatially resolved, and all the available information
about their geometry is obtained from analysis of variable lines. It is
well established that photoionization by the central radiation source
is the main source of ionization and excitation of the BLR gas. Indeed,
photoionization calculations, when applied to time-averaged spectra,
can reasonably explain most of the observed line ratios (for review and
references see Ferland, in these proceedings, and Netzer 1990). However,
such time-averaged calculations contain little information about the
spatial distribution of the gas.

Extensive monitoring campaigns, during the last decade, have produced
several high quality data sets. They include the time dependence of
the multi-wavelength continuum, as well as the change in line fluxes
and line profiles as a function of time (for a review see, Horne ---
these proceedings, Peterson 1993, Netzer \& Peterson 1997). Excellent
data sets are now available for half a dozen low luminosity AGNs. Less
complete (in terms of wavelength coverage) yet very detailed data sets,
are available on a dozen or so more sources.

Unfortunately, theoretical understanding lags behind and there are few,
if any, systematic attempts to produce complete BLR models that reproduce
the new light curves. Most recent studies focused on obtaining transfer
functions, and little effort has been devoted to reconstruct the physical
conditions in the gas. In particular, only one or two emission lines
have been considered while many more lines, and thus more information
and constraints, are available, at least in some data sets.

This work, as well as the more detailed results in Kaspi and Netzer
(1999), present an attempt to investigate one of the best data sets in a
new way. The goal is to reconstruct the observed light curves of as many
emission lines as possible in the Seyfert 1 galaxy NGC~5548. As shown
below, the observational constraints on the line intensity and line ratios
as a function of time, enable us to deduce the run of density, column
density and cloud distribution across the BLR in this source. Below we
demonstrate how the time dependent relative and absolute line intensities,
and their relationship to the variable continuum, leave little freedom
in modeling the BLR.

\section{Previous models}
\label{pre}

Previous attempts to model the BLR differ in the method used to
reconstruct the gas distribution and the assumptions made about the
BLR properties. We distinguish between direct and indirect methods.
Direct methods involve initial guess of the gas distribution and other
properties. These are later checked by calculating the predicted
emission line light curves, assuming the above properties and the
given variable continuum. Indirect methods attempt to obtain the gas
distribution by computing transfer functions for various emission
lines. This is somewhat ill-defined since it produces emissivity maps,
rather than mass-distribution maps. It therefore requires an additional
confirmation that the so-obtained emissivity maps are consistent with
photoionization calculations. While there were several attempts to produce
transfer functions for various lines, we are not aware of any successful
mapping that is consistent with full photoionization calculations (see
also Maoz 1994).

The first systematic attempt to reconstruct the BLR in NGC~5548 is by
Krolik et al. (1991). These authors used the Maximum Entropy Method to
reconstruct the transfer function and to model the BLR as a spherically
symmetric system of isotropically emitting clouds. The Krolik et al. BLR
is divided into two distinct zones: one emitting the high-ionization lines
(column density of $\sim$10$^{22}$ cm$^{-2}$ and ionization parameter
of 0.3) and the other emitting the low-ionization lines (column density
of $\sim$10$^{23}$ cm$^{-2}$ and ionization parameter of 0.1). Later,
O'Brien, Goad, \& Gondhalekar (1994) have combined photoionization
and reverberation calculations (\S~\ref{formalism}). Their study, and
also the one by P\'{e}rez, Robinson \& de la Funte (1992), focused on
the shape of the transfer function under different conditions and on
the line emissivity for a few generic models. They did not attempt any
detailed reconstruction of a specific data set.

Bottorff et al. (1997) presented a detailed kinematic model, combined with
photoionization calculations. This was applied to only {\it one line}
(C\,{\sc iv}$\lambda$1549\AA) in the spectrum of NGC~5548. Dumont,
Collin-Suffrin, \& Nazarova, (1998) modeled the BLR in NGC~5548 as a 3
zones region where the various component locations are determined by {\it
average} line-to-continuum lags. Much of their conclusions regarding the
required density and column density are based on the relative strength
of the Balmer lines. Finally, Goad \& Koratkar (1998) re-examined
the same NGC~5548 data set and deduced a strong radial dependence
($N \propto r^{-2}$ over the density range of $ 10^{11.3}-10^{10.0}$
cm$^{-3}$). Here again, the main assumption is of a simple two zone
model and the photoionization calculations are not applied to all the
lines. None of the above models presents a complete study or attempts
full recovery of all light curves. Hence, global consistency checks
are missing.

Our work relies heavily on the direct approach. We prefer to avoid, as
much as possible, ill-defined transfer functions and unreliable emissivity
distributions. Instead, we make a large number of initial assumptions
(``guesses'') and check them, one by one, against the complete data
set. This makes the numerical procedure more time-consuming but is more
robust because of the direct relationship between the assumed geometry
and the resulting light-curves.
 
\section{Formalism}
\label{formalism}

We follow the formalism presented in Kaspi \& Netzer (1999) which is first
described in Netzer (1990). The reader is referred to those references
for more information. In summary, we consider a spherical BLR consisting
of numerous small spherical clouds and a point-like ionizing source.
All important physical properties are represented by simple power-laws
in $r$, the distance from the central source. A possible justification
for this may be a radial dependent external pressure that determines the
cloud properties. The case presented here assumes isotropically emitting
clouds. We are fully aware of the potential complications due to this
assumption, in particular for lines like $L_{y \alpha}$.

The particle density in the model, $N(r)$ (assumed to be constant within
each cloud), is given by
\begin{equation}
N(r)\propto r^{-s} \ \ .
\label{den}
\end{equation}
The cloud column density, $N_{col}$, is computed by considering spherical
clouds, of radius $R_{c}(r)$. The mass of the individual clouds is
conserved, but it is not necessarily the same for all clouds, thus,
$R_{c}^{3}(r)N(r)=const$. The typical cloud column density is
\begin{equation}
N_{col}(r)\propto R_{c}(r)N(r)\propto r^{-2/3s} \ \ ,
\label{colden}
\end{equation}
and the geometrical cross-section is
\begin{equation}
A_{c}(r)\propto R_{c}^{2}(r)\propto r^{2/3s} \ \ .
\end{equation}
The number density of such clouds per unit volume is 
\begin{equation}
n_{c}(r)\propto r^{-p} \ \ .
\end{equation}
The clouds are illuminated by a central source whose ionizing luminosity,
$L(t)$, varies in time. Designating $\epsilon_{l}(r,L)$ as the flux
emitted by the cloud in a certain emission-line -- $l$ per unit projected
surface area ($erg \ s^{-1} \ cm^{-2}$), we find the following relation
for a single cloud emission:
\begin{equation}
j_{c,l}(r)=A_{c}(r)\epsilon_{l}(r,L) \ \ .
\end{equation}
Assuming the system of clouds extends from $r_{in}$ to $r_{out}$ we 
integrate over $r$ to obtain cumulative line fluxes:
\begin{equation}
E_{l}\propto \int_{r_{in}}^{r_{out}}n_{c}(r)j_{c,l}(r)r^{2}dr \ \ .
\label{El}
\end{equation}
Having determined the properties of the emission line clouds, and having
assumed a spectral energy distribution (SED) for the ionizing source,
we now calculate $\epsilon_{l}(r,L)$ using a photoionization code and
follow the formalism to obtain $E_l$.

We also consider the changes of $L$, and possibly also the SED, in
time. We take this into account by calculating $\epsilon_{l}(r,L(t))$ for
the entire range of continuum luminosity applicable to the source under
discussion. The calculated line fluxes are the results of integrating
Eq.~\ref{El}, using, at each radius, the relevant ionizing flux, i.e. the
one obtained with the ionizing luminosity $L(t-r/c)$.

A model is specified by the source luminosity and SED, the radial
parameter $s$, and the normalization of the various free parameters.
These include $p$, $r_{in}, r_{out}$ and the density and column density
at a fiducial distance which we take to be one light-day. The comparison
with observations further requires the normalization of the total line
fluxes and hence the integrated number of clouds (an alternative way
of presenting this normalization is by defining a radial dependent
covering fraction).

We have calculated a large grids of photoionization models covering the
entire range of density, column density, and incident flux applicable for
this source. The calculations were performed using ION97, the 1997 version
of the code $ION$ (see Netzer 1996, and references therein). There are
several limitations for such codes which should be considered. Most
important (and crucial for any BLR model) is the transfer of the optically
thick lines. This is treated with a simple, local escape probability
method which has long been suspected to be inadequate for the Balmer
lines (see Netzer 1990). The problem is not yet solved and is common to
most detailed photoionization models similar to $ION$, like $Cloudy$ by
G. Ferland. There is no simple solution for this problem and we prefer,
at this stage, not to consider Blamer lines in this work. There is a
similar problem for several other low ionization lines, like Mg\,{\sc
ii}$\lambda$2798\AA\ and the FeII lines. On the other hand, the transfer
of lines like Ly$\alpha$ and C\,{\sc iv}$\lambda$1549\AA\ is much better
understood.

\section{A model for NGC 5548}

The Seyfert 1 galaxy NGC 5548 is one of the best studied AGN. It was
monitored, for 8 months, in the optical-UV, in 1989, and was also the
subject of an intensive optical spectroscopic monitoring for 8 years
(Peterson et al, 1999). Several shorter monitoring campaigns, each with
a duration of several months, took place in various wavelength bands.
This makes NGC~5548 an excellent choice for testing our model. In this
study we have concentrated on the $International \ Ultraviolet \ Explorer
\ (IUE)$ 1989 campaign (Clavel et al. 1991) and used the resulting
light curves of the UV continuum (at $\lambda$1337\AA) and the following
emission lines: Ly$\alpha\lambda$1216\AA , C\,{\sc iv}$\lambda$1549\AA
, C\,{\sc iii}]$\lambda$1909\AA , He\,{\sc ii}$\lambda$1640\AA , and
Mg\,{\sc ii}$\lambda$2798\AA . (All diagram in this paper show light
curves of only the first three lines.)

Several additional observational constraints have been considered:
\begin{enumerate} 
\item 
The $IUE$ spectra clearly show that the C\,{\sc iii}]$\lambda$1909\AA
\ line is blended with Si$\lambda$1895\AA. Clavel et al. (1991) have
measured the combined flux of both lines, and listed it as C\,{\sc
iii}]$\lambda$1909\AA . Hence, our calculated flux for both lines will
be combined (see Kaspi and Netzer 1999 for more details regarding the
use of this ratio as a density diagnostics).
\item
The observed SED of NGC~5548 is reviewed by Dumont et al. (1998). We
found a strong dependence of our results on the ratio of the UV to
X-ray continuum flux. We have chosen a typical Seyfert 1 SED with
$\alpha_{ox}=1.06$ similar to the one measured for NGC 5548.
\item
Using the observed $\lambda$1337\AA \ continuum flux, an assumed
cosmology ($H_0$= 75 km\,s$^{-1}$Mpc$^{-1}$ and $q_0$=0.5), and assumed
SED, we have estimated a time averaged ionizing luminosity of
$10^{44}$ erg s$^{-1}$. During the $IUE$ campaign, the UV continuum
flux varied by a factor of $\sim 4.5 $, hence we choose our grid to
cover the ionizing luminosity range of $10^{43.6}$ to $10^{44.3}$ erg
s$^{-1}$.
\item
Based on reverberation mapping results, we have defined our grid of
distances to cover the range of 1 to 100 light days.
\end{enumerate} 

Model calculations proceed in two stage. First we produce a two
dimensional grid of $\epsilon_{l}(r,L(t))$ for several values of
the parameter $s$. We have considered $s$=1, 1.5, and 2, with the
additional normalizations of the density at 10 light day ($N(r$=10))
in the range $10^{9.5}$ to $10^{11.5}$ cm$^{-3}$, and column densities
($N_{col}(r$=10)) of $10^{21}$ to $10^{23}$ cm$^{-2}$. In the second stage
we calculate theoretical light curves by integrating over Eq.~\ref{El}
with the given $s$, the chosen $r_{in}$ and $r_{out}$, and several values
of the parameter $p$. We have examined models with $p$=1, 1.5, and 2 for
each grid. In each model we vary both $r_{in}$ and $r_{out}$ to minimize
$\chi^{2}$. This score is calculated by comparing the theoretical and
observed lines fluxes for {\it all chosen lines}. The smallest $\chi^{2}$
determines our best parameters.

\subsection{Detailed examples}
\label{example}

To illustrate our model, we present the results of two sets of
calculations. First we consider a model with $s$=2 and $p$=1.5 (This
particular choice results in $N_{col}(r)\propto r^{-4/3}$). We chose
$N_{col}(r$=10)=10$^{22.67}$ cm$^{-2}$ and study the full density range.
We note again that is referred to here as {\it a single model} covers in
fact a large range of density and column density. For example, choosing
$N(r$=10)=10$^{10.5}$ cm$^{-3}$ means a full density range of 10$^{9\,
- \, 12}$ cm$^{-3}$ and a full column density range of 10$^{21.33\, -
\, 24}$ cm$^{-2}$.

\begin{figure}[t]
\plotfiddle{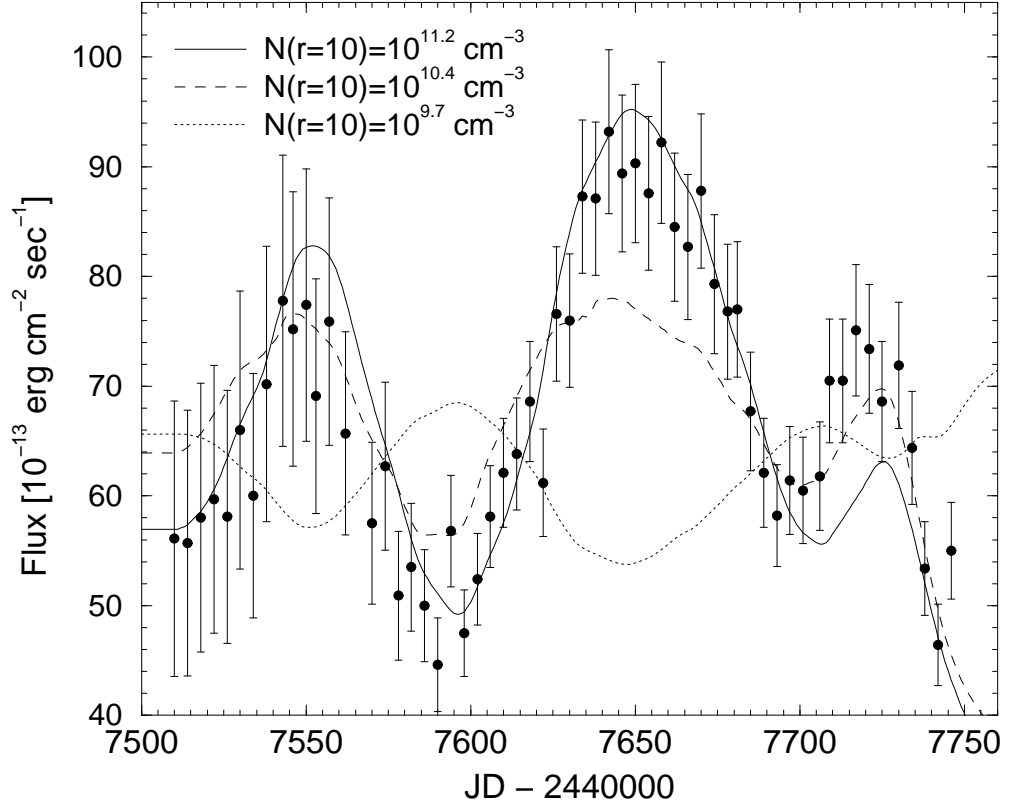}{5.0cm}{0}{80}{80}{-218}{-190}
\vspace{5.2cm}
\caption{Ly$\alpha$ light curves resulted from a model with $s=2$,
$r_{in}$=3 ld, $r_{out}$=25 ld $N_{col}(r$=10)=10$^{22.67}$ cm$^{-2}$,
and various values of $N(r$=10) as marked. The observed light curves
are from Clavel et al. (1991).}
\label{la_m2}
\vspace{-0.2cm}
\end{figure}

Examining this case we note some obvious features. First,
for $N(r$=10)${\mathrel{\raise.3ex\hbox{$<$}\mkern-14mu
\lower0.6ex\hbox{$\sim$}}}$ 10$^{10.2}$ cm$^{-3}$ the response of the
modeled emission lines is reversed, i.e., increasing continuum flux
results in decreasing line flux. This is demonstrated in Fig.~\ref{la_m2}
(check the doted line in the diagram). This is a clear sign of optically
thin material in the inner BLR. It is caused by a combination of large
incident fluxes and small densities (i.e. large ionization parameters)
with relatively small column density. Such models are in poor agreement
with observations and smaller ionization parameters and/or larger columns
must be considered.

\begin{figure}
\plotone{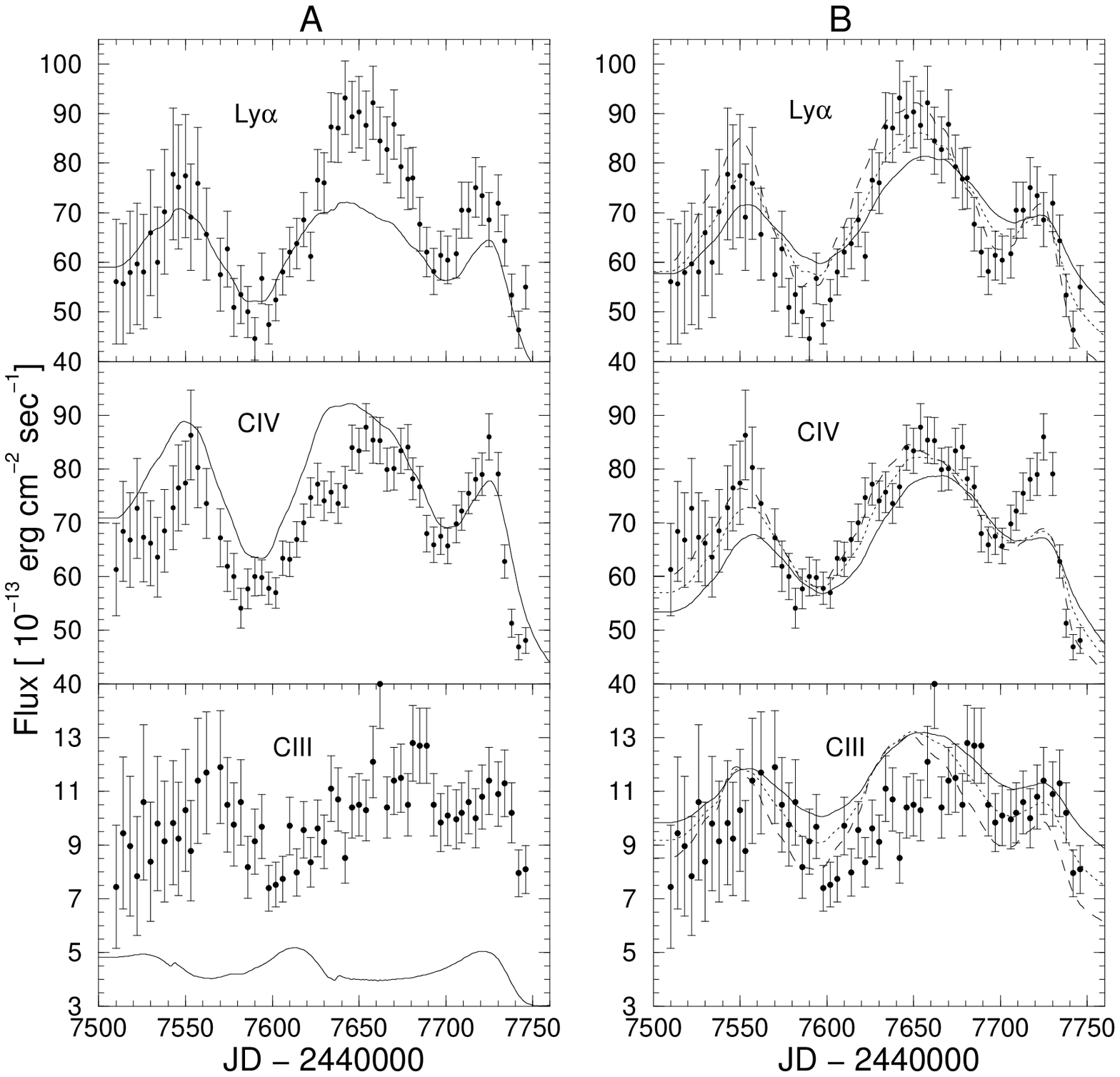}
\caption{Simulated light-curves compared with observations:
\protect\vspace{0.3cm}
\newline
{\bf A.} A model with $s$=2, $p$=1.5,
$N(r$=10)=10$^{10.4}$ cm$^{-3}$, $N_{col}(r$=10)= 10$^{22.67}$ 
cm$^{-2}$, $r_{in}$=3 ld, and $r_{out}$=25 ld. The bad disagreement
between the observed and theoretical light curve is typical to all
$s=2$ models.
\protect\vspace{0.3cm}
\newline
{\bf B.} A model with $s$=1, $N(r$=10)=10$^{10.4}$ cm$^{-3}$,
$N_{col}(r$=10)=10$^{23.33}$ cm$^{-2}$, $r_{in}$=3 ld, and
$r_{out}$=100 ld. Models for different values of $p$ are presented:
solid line -- $p$=1 ; dotted line -- $p$=1.5 ; dashed line -- $p$=2.}
\label{test3}
\end{figure}

The problem can be cured by raising the density to
$N(r$=10)${\mathrel{\raise.3ex\hbox{$>$}\mkern-14mu
\lower0.6ex\hbox{$\sim$}}} $10$^{10.2}$ cm$^{-3}$. The resulting
light curves is in much better agreement with the observations
(Fig.~\ref{la_m2}: dashed and solid lines). The reversed response in
Ly$\alpha$ has disappeared and the theoretical line light curves nicely
fit the observed ones with $r_{in}$=3 ld and $r_{out}$=25 ld assuming
$N(r$=10)=10$^{10.4}$ cm$^{-3}$. However, applying this geometry
to the other emission lines, we get unsatisfactory results. This is
illustrated in Fig.~\ref{test3}A (solid lines) that show the improved
fit to the Ly$\alpha$, and C\,{\sc iv} data and the very poor agreement
for C\,{\sc iii}].

The above example demonstrate the difficulty in fitting, simultaneously,
line responses and line ratios. The example illustrates why a particular
radial dependence, like the one used here ($s=2$), cannot adequately fit
the observed variable emission line spectrum of NGC~5548. Our simulations
show that no normalization or a choice of $p$ can cure this problem.

In the second example $s$=1 with the following normalization:
$N(r$=10)= 10$^{10.4}$ cm$^{-3}$, $N_{col}(r$=10)=10$^{23.33}$ cm$^{-2}$,
$r_{in}$=3 ld and $r_{out}$=100 ld. The light curves are presented in
Fig.~\ref{test3}B. This set of models give much better fit to the data
(note in particular the C\,{\sc iii}] light-curve). Comparing both
models demonstrate the usefulness of this approach in constraining the
parameter space.

\subsection{General trends}

We have checked a large range of model parameters and comment, below, on
several of the more obvious trends.

We first note that as the column density grows from
$N_{col}(r$=10)=10$^{21}$ to $10^{23}$ cm$^{-2}$, the agreement
with the observations is much better. Thus, models with
$N_{col}(r$=10)=10$^{21}$ cm$^{-2}$ do not fit the observations
while $N_{col}(r$=10)${\mathrel{\raise.3ex\hbox{$>$}\mkern-14mu
\lower0.6ex\hbox{$\sim$}}}$10$^{22}$ cm$^{-2}$ already results in
a reasonable agreement. Photoionization codes like the one used
here, are limited to column density less than about $10^{25}~cm^{-2}$
(i.e. Compton thin clouds). Hence we are unable to put an upper limit
on the column density.

Using models with $s$=1.5 and $s$=1, we can obtain
lower and upper limits on the gas density. Models
with $N(r$=10)${\mathrel{\raise.3ex\hbox{$<$}\mkern-14mu
\lower0.6ex\hbox{$\sim$}}}$10$^{9.5}$ cm$^{-3}$ do not fit the
observation (in the $s$=1.5 models there is a reverse response of the
lines and in the $s$=1 models the line ratios do not agree with the
observations). For $N(r$=10)${\mathrel{\raise.3ex\hbox{$>$}\mkern-14mu
\lower0.6ex\hbox{$\sim$}}}$10$^{11}$ cm$^{-3}$, the line ratios for
both value of $s$ do not agree with the observations. Hence we find
the BLR density of NGC~5548, in our model, to be in the range of
10$^{11}$$>$$N(r$=10)$>10^{9.5}$ cm$^{-3}$.

A general trend for all values of $s$ is that as $p$ increases from 1
to 2 (i.e., more weight is given to clouds closer to the central source)
the amplitude of the modeled light curves is in better agreement with the
observations and so are most line ratios. Hence, models with higher $p$
are preferred.

A common problem for all models is the weak Mg\,{\sc ii}$\lambda$2798\AA\
line. While we do not have a complete explanation for this, we suspect
it is due to one of two reasons: either the transfer of this line is
inaccurate, similar to the case of the Balmer lines, or else it is caused
by the thousands of highly broadened Fe\,{\sc ii} lines, in that part
of the spectrum, that make the measured line intensity highly uncertain
(see for example the discussion in Wills, Wills and Netzer, 1985; Maoz
et al. 1993) (a third possibility of enhanced metallicity is discussed
in Kaspi and Netzer 1999). Our $\chi^{2}$ evaluation does not include
the Mg\,{\sc ii} line. This is a definite failure of the model.

Another common trend is the improvement of the $\chi^2$ score with
increasing $r_{out}$. This is most noticeable as $r_{out}$ increases
from 50 to 100 light days.

Considering all the above trends and limitations, and using the $\chi^{2}$
score, we find that models with $s$=1 best fit the observed spectra and
models with $s=1.5$ give somewhat inferior fits. An example of one of
our best models is shown in Fig.~\ref{test3}B. In this model the reduced
$\chi^{2}$ score for the four lines is 4.5 for the $p$=1 model, 3.1
for the $p$=1.5 model, and 2.2 for the $p$=2 model. The total covering
factors found for these models are 0.25, 0.28. and 0.30, respectively.

\section{Discussion}
\label{discussion}

Our direct method allow us to investigate, in a critical way, a large
variety of BLR models. Unlike indirect methods that are based on
transfer function of individual lines, and make no use of their relative
or absolute intensity, we are able to introduce many more observational
constraints. The $\chi^2$ minimization applied to 4 emission lines,
at {\it all times}, enable us to choose among various models and to rule
out cases of unsuitable density, column density and covering fraction. In
particular we were able to show that:
\begin{enumerate}
\item
There is a narrow range of density and density dependence that fit the
observed light cures. Using our parameters we find that $s$ is in the
range of 1--1.5 and the largest density (at one light day) is about
10$^{12.5}$ cm$^{-3}$. The simulations rule out steep density laws like
$s=2$. This is in disagreement with the results of Goad \& Koratkar
(1998) despite of the fact that their range of acceptable densities
($N\sim$10$^{11.3}$ cm$^{-3}$ in the inner BLR and $N\sim$10$^{10.0}$
cm$^{-3}$ in the outer BLR) is similar to ours and their deduced lower
limit on the column density is also in agreement with ours. We suspect
that Goad \& Koratkar's conclusion about the good fit of the $s=2$ case
is related to their use of mean time lags rather than compared with our
detailed fit to the light curves.
\item
Detailed emission line variability can be used to put useful constraints
on the column density of clouds across the BLR. Realistic models require
large enough columns to avoid optically thin clouds at small distances.
\item
Simple two or three zone models contain too little information and hence
cannot constrain, accurately enough, the physical conditions in the
gas. For example Dumont et al. (1998) have used a three zone model and
reached several conclusions based on time-averaged properties. They noted
three problems arising from their modeling: an energy budget problem,
a line ratio problem, and a line variation problem. The first two are
most probably related to the Balmer line intensity and, as explained,
we suspect this to be a general limitation of current photoionization
models. Regarding the line intensity, we find good agreement for both high
and low ionization lines and suspect that even a three-zone model is
highly simplified for the purpose of realistic reconstruction of the BLR.
\item
Several recent ideas about the BLR can be tested against real observations
by using an approach similar to ours. `Locally Optimally-emitting Clouds'
(LOCs) models (Baldwin et al. 1995, Korista et al. 1997) have been
suggested to explain the broad line spectrum of AGNs. The models assume
that there are clouds with a range of density and column density at each
distance. LOCs must be put into a real test by checking whether they
result in light curves that are in better agreement with the observations,
compared with the simpler models assumed here.

Alexander \& Netzer (1997) have suggested that the BLR clouds may
be bloated stars (BSs) with extended envelopes. In their work they
fitted the emission-line intensities, profiles and variability to mean
observed properties of AGNs. One of the conclusion is that the density
at the external edge of the BSs (the part emitting the lines) falls off
like $r^{-1.5}$, and the number density of BSs falls off like $r^{-2}$.
These two trends are in good agreement with our preferred values of $s$
and $p$. While the model is consistent with mean time-lags of intermediate
luminosity AGNs, it remains to be seen whether it can fit, in detail,
the time dependent spectrum of objects like NGC~5548.
\end{enumerate}

There are obvious limitations and several ways to improve our models.
First, different abundances ought to be considered. Second, line
beaming (anisotropy in the emission line radiation pattern) must
be considered. Unfortunately, similar to the Balmer line problem,
the radiation pattern cannot be accurately calculated in present-day
escape-probability based codes. Multi-component models, with a range of
density and column density at each radius (not necessarily similar to
the LOC distribution), must be tested too. Finally, a variable shape SED
(Romano and Peterson 1998) is a likely possibility that may affect the
outcome of such models. Some of these additional factors are addressed
in Kaspi and Netzer (1999). Others must await future work.

\acknowledgments

This work is supported by a special grant from the Israel Science
Foundation.

\end{document}